\documentclass[journal,10pt,letterpaper,final,twocolumn]{IEEEtran}%
\usepackage{amsmath}
\usepackage{amsfonts}
\usepackage{cite}
\usepackage{amssymb}
\usepackage{mathrsfs}
\usepackage{ragged2e}
\usepackage{graphicx}
\usepackage[usenames,dvipsnames]{color}
\usepackage{epstopdf}
\usepackage[all]{xy}
\usepackage[]{mcode}
\usepackage{booktabs}
\usepackage{supertabular}
\usepackage{subcaption}
\usepackage{multirow}
\usepackage{array}

\providecommand{\U}[1]{\protect\rule{.1in}{.1in}}
\pdfoutput=1
\providecommand{\U}[1]{\protect\rule{.1in}{.1in}}

\newenvironment{problem}[1][Problem]{\begin{trivlist}
\item[\hskip \labelsep {\bfseries #1}]}{\end{trivlist}}

\newenvironment{definition}[1][Definition]{\begin{trivlist}
\item[\hskip \labelsep {\bfseries #1}]}{\end{trivlist}}

\newcommand{\qed}{\nobreak \ifvmode \relax \else
      \ifdim\lastskip<1.5em \hskip-\lastskip
      \hskip1.5em plus0em minus0.5em \fi \nobreak
      \vrule height0.75em width0.5em depth0.25em\fi}
      
\begin{document}

\title{Prospect Theory for Human-Centric Communications}
\author{Kevin Luo, Shuping Dang, \textit{Member, IEEE}, Basem Shihada, \textit{Senior Member, IEEE}, and \\Mohamed-Slim Alouini, \textit{Fellow, IEEE}
  \thanks{K. Luo is with Graduate School of Economics, Kobe University, Kobe 657-8501, Japan (e-mail: kevin.luo@stu.kobe-u.ac.jp).
  
  S. Dang, B. Shihada, and M.-S. Alouini are with Computer, Electrical and Mathematical Science and Engineering Division, King Abdullah University of Science and Technology (KAUST), 
Thuwal 23955-6900, Kingdom of Saudi Arabia (e-mail: \{shuping.dang, basem.shihada, slim.alouini\}@kaust.edu.sa).}}

\maketitle

\begin{abstract}
Entering the 5G/6G era, the core concept of human-centric communications has intensified the search effort into analytical frameworks for integrating technological and non-technological domains. Among non-technological domains, human behavioral, psychological, and socio-economic contexts are widely considered as indispensable elements for characterizing user experience (UE). In this study, we introduce the prospect theory as a promising methodology for modeling UE and perceptual measurements for human-centric communications. As the founding pillar of behavioral economics, the prospect theory proposes the non-linear quantity and probability perception of human psychology, which extends to five fundamental behavioral attributes that have profound implications for diverse disciplines. By expatiating on the prospect theoretic framework, we aim to provide a guideline for developing human-centric communications and articulate a novel interdisciplinary research area for further investigation.  
\end{abstract}


\section{Introduction}
\IEEEPARstart{T}{elecommunication} research so far has tended to become technology-centric and focused excessively on quality of service (QoS) metrics. On the other hand, due to the rapid emergence of end-user controllable and programmable devices in communication networks, quality of experience (QoE) and user experience (UE) have increasingly attracted researchers' attention in both academia and industry in recent years \cite{kilkki2008quality,6679038,8447199,finley2017does,boz2019mobile}.  Furthermore, the development of human-centric communications in the 6G network requires not only technological factors but also UE to be considered when modeling, analyzing, and optimizing communication systems \cite{Dang2019FromAH}. This paradigm shift necessitates the interdisciplinary collaboration among telecommunications, economics, and psychology. 

The first attempt to propose an ecosystem incorporating multiple stakeholders and end users is presented in \cite{kilkki2008quality}, based on which several advanced versions of ecosystems are proposed in \cite{6178834,dong2014quality,reichl2010charging,8701705}. However, due to the absence of a quantitative basis, existing ecosystems can hardly be utilized to carry out performance analysis for communication systems. To facilitate quantitative analysis, a standard approach is to measure UE via utility functions, such as logarithmic, sigmoid, and exponential functions \cite{reichl2013logarithmic,6891176,5430142}. However, most QoE measurements are confined to simple quantitative relationships between QoS and QoE metrics. So far, no analytical framework has been proven to be generic and theoretically practicable for modeling UE in a human-centric manner.

In the area of economics, researchers resort to the prospect theory for accurately predicting decision-making behaviors of human beings \cite{kahneman1979prospect,tversky1992advances}. The prospect theory is a Nobel prize winning theory and the founding pillar of behavioral economics. It is widely perceived as the most satisfactory descriptive theory of quantity perception currently available. Since its proposal in 1979, this theory has been extensively applied in pricing strategy, labor supply, tax policy making, and finance related topics. Due to its generality and versatility, the prospect theory is also introduced as a nexus to connect the disciplines of telecommunications, economics, and psychology. In \cite{6747282}, the user behavior interference in networking protocols is modeled and analyzed according to the prospect theory. Specific resource allocation strategies using the prospect theory are investigated in \cite{7560647}. Data pricing problems relying on the prospect theory for licensed and unlicensed communications are investigated in \cite{7869340} and \cite{7294655,8746552}, respectively.  The prospect theory has also been applied to provide secure protection mechanisms for communication systems by formulating dynamic defense games \cite{7842178,7835168}. Incorporating game theory, the prospect theory can also help communication systems equip with better anti-jamming and random access functions \cite{7036897,6310922}. A generic but simplistic prospect theoretic analytical framework of UE is proposed in \cite{7346204}. Even though instructive and creative, the inappropriate and oversimplified modeling and assumptions in \cite{7346204} result in a huge mismatch between the quantitative UE and resource utilization. In these interdisciplinary applications, however, the prospect theory serves only as a replacement for the utility or probability functions formulated in specific problems (e.g., game theoretic communications). This, to some extent, undervalues the prospect theory's implications and application aspects for human-centric communications.

In light of this, we summarize the contributions of this paper as follows:
\begin{itemize}
\item We outline five essential attributes of the prospect theory considering user psychology that should be taken into consideration for wireless system modeling.
\item We construct a comprehensive analytical framework for modeling UE and perceptual measurements for human-centric communications.
\item We detail how modeling and analysis of communication systems will be reshaped under the novel analytical framework with several exemplary cases. The proposed analytical framework can be directly applied to carry out performance analysis for most communication systems when non-technological factors are taken into consideration and can be easily tailored to fit a broader range of communication applications. 
\end{itemize}
By the contributions given in this paper, we aim to provide a guideline for improving communication services and articulate a new interdisciplinary research area for further investigation. 

The remainder of this paper is as follows. In Section \ref{fpt}, we present the fundamentals of the prospect theory, including its economic background, concepts, intuitions, and five key attributes of human psychology. Based on the five key attributes, we formulate the prospect theoretic analytical framework in Section \ref{ptbaf}. In Section \ref{cs}, we utilize several case studies to demonstrate how the analytical framework can be used to evaluate the subjective perception of communication systems. We outline the challenges and promising research directions for further investigations in Section \ref{fird} and conclude the paper in Section \ref{c}.

\section{Fundamentals of the Prospect Theory}\label{fpt}

As we are entering the 5G/6G era, the development of human-centric communications is becoming increasingly important but faces tremendous challenges. The paradigm shift from technology-centric to human-centric applications necessitates the interdisciplinary collaboration among telecommunications, economics, and psychology. In the following subsections, we explain step by step how this interdisciplinary collaboration is enabled by the prospect theory.

\subsection{Existing Problems with Classic Telecommunication Research}
Researchers employ different approaches to establish linkages between QoS (e.g., bandwidth and loss rate) and QoE. Perceptual measurements such as the mean opinion score (MOS) and the pseudo-subjective quality assessment (PSQA) have been developed to quantify user experience \cite{streijl2016mean,5070785}. There are also multi-dimensional evaluation systems designed to assess network performance from end-user perspectives \cite{6178834}. Although the QoE research is gaining strong momentum in recent years, the literature remains scattered, inconsistent, and excessively technology-centric \cite{kilkki2008quality,reichl2010charging,de2010proposed}. So far, no analytical framework has been proven sufficiently generic and theoretically practicable for modeling UE in a user-oriented manner. 

Classic communication research focuses on technology-centric network optimization and bypasses the model of UE. In such settings, researchers have actually made an implicit assumption: the optimization process maximizes both network capacity and user's subjective utility simultaneously \cite{barakovic2013survey}. Conventionally, for $n$ mutually exclusive outcomes $y_i$ with occurrence probability $p_i$, the subjective utility can be considered as a linear function of probabilistic outcomes: $U=\sum_{i=1}^{n}p_iy_i$, given $\sum_{i=1}^{n}p_i=1$. However, the linear utility function is difficult to reconcile with human psychology due to its unrealistic attributes:
\begin{itemize}
\item Subjective utility is solely determined by the state of the final outcome.
\item Marginal utility is constant.
\item Quantity perception over gains and losses is symmetric.
\item There is indifference between objective and perceived probabilities.
\end{itemize}
Under the erroneous characterization of human psychology, the optimization process does not necessarily guarantee optimal UE, and the derived theories would be of limited practical usefulness. As an applied discipline, communications become human-centric only if UE can be properly introduced and optimized in the modeling process, which should be the very foundation of communication science. Overall, there is an urgent need to incorporate UE into the modeling of telecommunication theories.

\subsection{Prospect Theory}
In the past decades, the rapid development of behavioral economics has greatly enriched our understanding of human psychology and proposed promising analytical frameworks for modeling UE. As the founding pillar of behavioral economics, the prospect theory is widely perceived as the most satisfactory descriptive theory of quantity perception and decision making currently available \cite{tversky1992advances,kahneman1979prospect}. The prospect theory is also a Nobel Prize winning theory, and the research article \cite{kahneman1979prospect} is the second most cited paper in economics. The prospect theory provides a well-established theoretical framework and mathematical tools for modeling real-life UE, and its gist is that the human perception of quantity and probability are non-linear.

\begin{figure*}[!t]
    \centering
    \begin{subfigure}[t]{0.5\textwidth}
        \centering
        \includegraphics[width=3.5in]{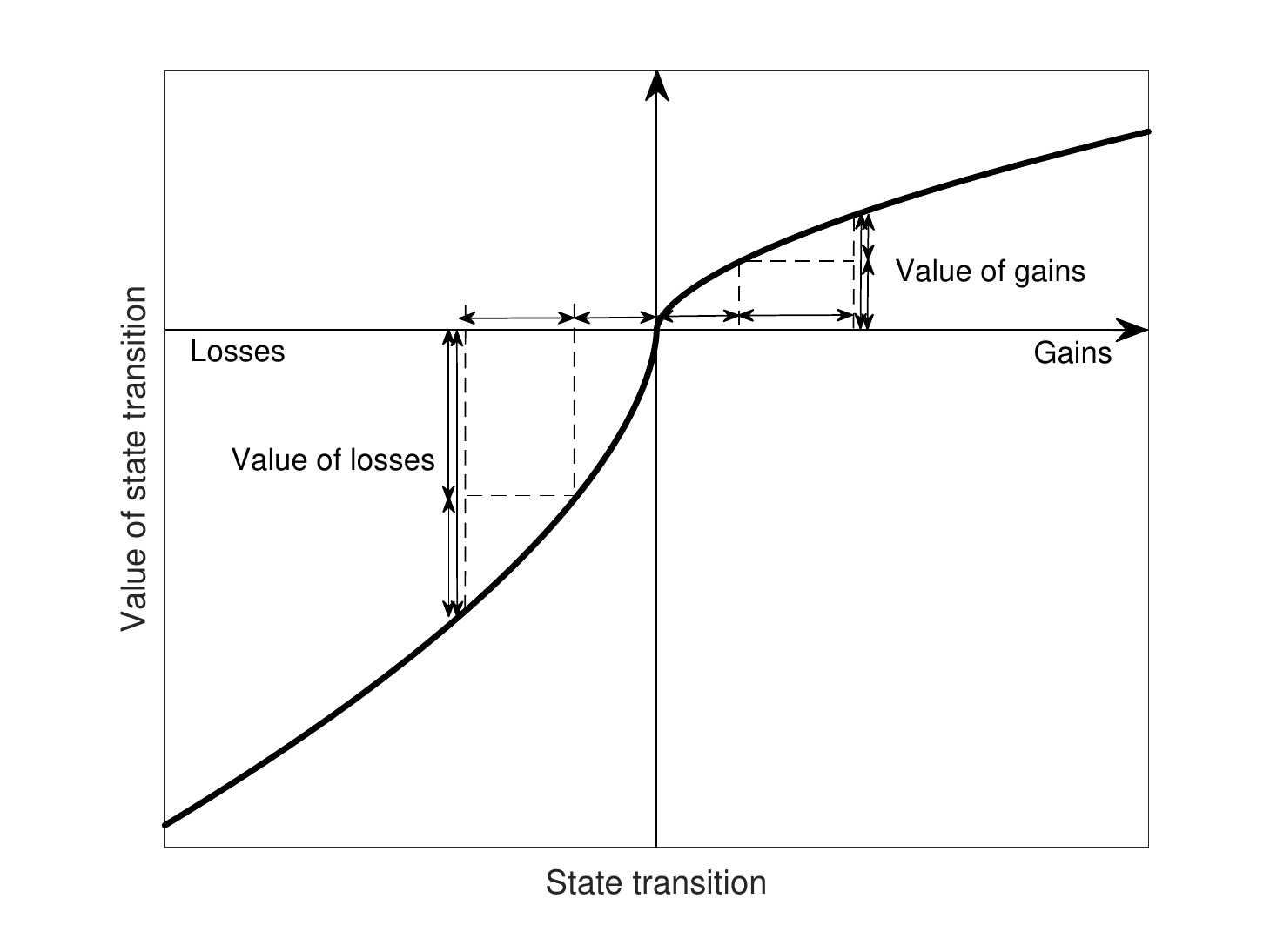}
        \caption{}
    \end{subfigure}%
~
    \begin{subfigure}[t]{0.5\textwidth}
        \centering
        \includegraphics[width=3.5in]{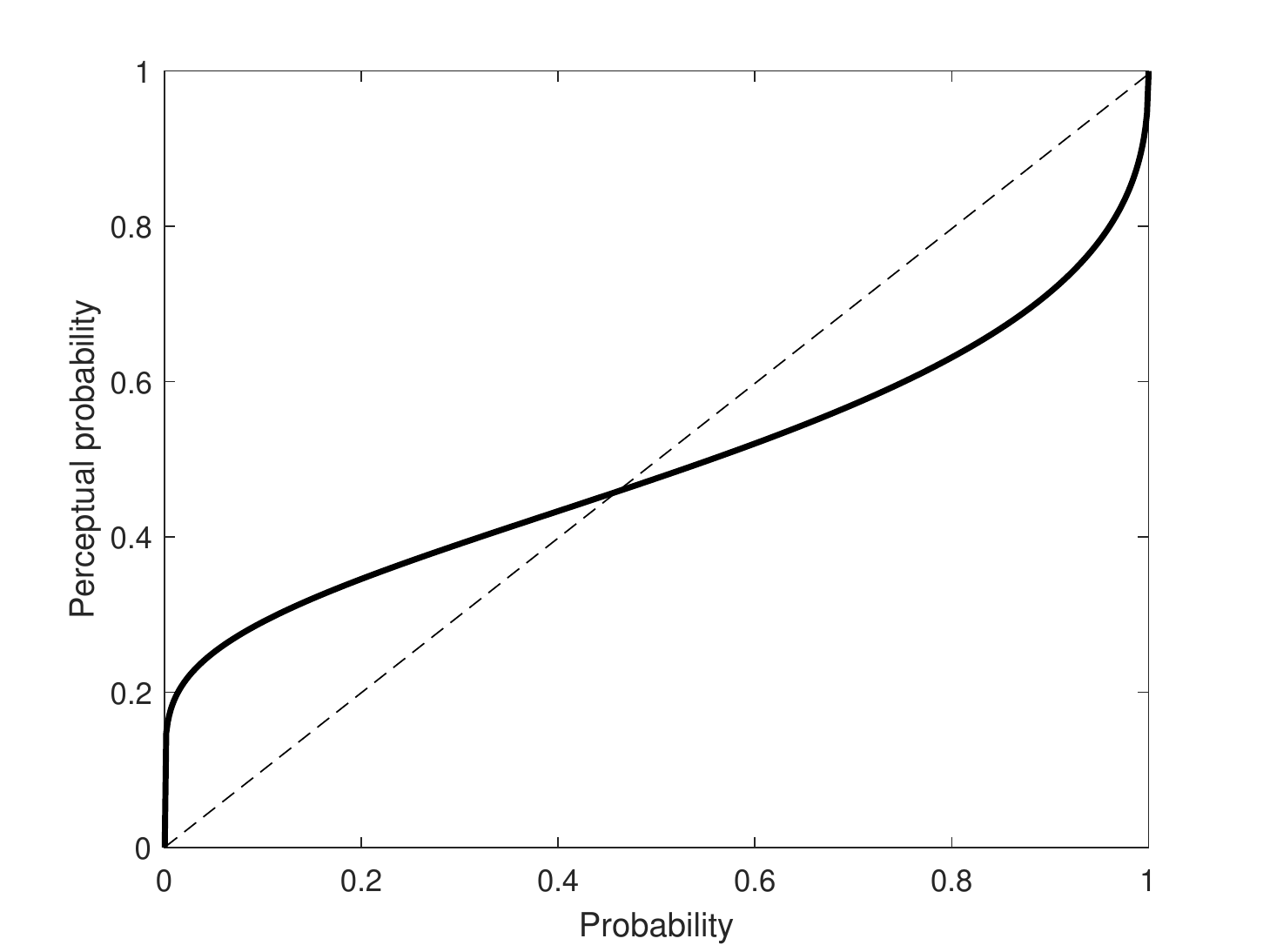}
        \caption{}
    \end{subfigure}
    \caption{(a) Perceptual quantity under the prospect theory; (b) Perceptual probability under the prospect theory \cite{dhami2016foundations}.}
    \label{lizi}
\end{figure*}

In the current literature, most QoE measurements for telecommunication studies are confined to quantity perception of discrete states, i.e., quantitative relationships between UE and QoS \cite{5430142,reichl2013logarithmic}. As a major advantage, the prospect theory characterizes two indispensable dimensions of UE, i.e., quantity and probability, and can help to model and analyze UE in the continuous state. Under the prospect theory, the perception of quantity and probability perception can be expressed in Fig. \ref{lizi}. This figure is believed to be the most important pictorial illustration of the prospect theory and can help with the mapping from QoS to QoE for telecommunication studies. A value function and a probability weighting function are employed to model the human perception, which can be characterized by the two-part functional form of \cite{tversky1992advances} and the Prelec function given in \cite{prelec1998probability}, respectively.

\subsection{Five Key Attributes}
In this subsection, we introduce the features of the prospect theory in shaping human psychological foundations and discuss the insights into modeling UE. According to Fig. \ref{lizi}, we can observe five important features of human psychology, captured by the prospect theory as follows.

\subsubsection{Reference dependence}
The prospect theory states that individuals perceived value through changes (i.e., quantity deviations from a reference point) instead of states. There is always a reference point in each dimension of quantity perception. This can be explained by the following thought experiment. Communication operators upgrade Jack's mobile network from 3G to 4G but downgrade Jim's from 5G to 4G. Jack and James now have the same network capacity but markedly different quantity perception (gains versus losses), due to reference dependence. This characterization is consistent with the advanced notion of QoE: users' perceived value can be either positive (delight) or negative (annoyance) \cite{le2012qualinet}. As a fundamental trait in human psychology, reference dependence casts doubts on optimization approaches that focus merely on the level of final outcomes, which has also attracted research attention in recent QoE literature \cite{5430142}.
\subsubsection{Diminishing marginal utility}
Individuals have diminishing sensitivity towards the scale of changes. This means that the subjective difference between the data packages of 50 Mb and 100 Mb is more salient than that between the data packages of 950 Mb and 1000 Mb. Diminishing marginal utility, also known as diminishing sensitivity and utility curvature, is a widely recognized psychological feature shared by living creatures. Under proper assumptions (e.g., exponential or logarithmic utility functions \cite{5430142,reichl2013logarithmic}), the concept of diminishing sensitivity can be readily incorporated into the modeling process to improve the practical significance of telecommunication theories.
\subsubsection{Loss aversion}
Individuals are more sensitive to losses than equivalent gains. Ample experimental evidence suggests that the magnitude of loss aversion is context-dependent, but losses are generally twice as significant as equivalent gains in quantity perception \cite{dhami2016foundations}. The phenomenon of loss aversion is captured by the value function  in Fig. \ref{lizi}(a), where quantity perception is steeper in the domain of losses. This asymmetry calls for a reconsideration of conventional approaches that weight gains and losses equally.
\subsubsection{Asymmetric risk attitudes}
Under diminishing marginal utility and loss aversion, individuals are risk seeking in the domain of losses but risk averse in the domain of gains. A prime example is the behavior of compulsive gambling: money-winning gamblers tend to play safely and prefer low-risk-low-return options, whereas money-losing gamblers prefer high-risk-high-return options and expect a grand slam home run. Asymmetric risk attitudes indicate that users have different risk preferences for improvement and deterioration in communication services, which calls for revisions of risk modeling in the QoE research. 
\subsubsection{Probability distortion}
Individuals tend to overweight small probabilities and underweight moderate and high probabilities. Consider the thought experiment as follows:
\begin{problem}
\textbf{1} \textit{How much would you pay to reduce the dropping probability from 5\% to 0\%.}
\end{problem}
\begin{problem}
\textbf{2} \textit{How much would you pay to reduce the dropping probability from 55\% to 50\%.}
\end{problem}

The two problems seem identical because they are about the perceived value of 5\% risk of dropping call. However, the great majority of people in real life are willing to pay much higher in \textbf{Problem 1} than in \textbf{Problem 2}. This phenomenon suggests a non-linear subjective probability weighting, in contrast to the linear probability weighting that equates objective probability with perceptual probability\footnote{Similarly, one can consider the acceptable amount of compensation if the dropping probability increases by 5\%. In most cases, people would not allow an increase of dropping probability from 0\% to 5\% but might be willing to negotiate over the compensation if the dropping probability increases from 50\% to 55\%. Again, the contrast between the increase and decrease of dropping probability is due to loss aversion, i.e., loss is subjectively more significant than equivalent gain.}. In probability perception, probability changes such as 0-5\% and 95-100\% are subjectively more salient than other changes of the same magnitude, since they are qualitative changes between non-existence, probabilistic outcome, and certainty. Extensive experimental evidence indicates that people are rather sensitive to the edges of probability interval [0,1], as documented by the inverse S-shaped probability weighting in Fig. \ref{lizi}. 

\subsection{Summary}
Under the prospect theory, the five psychological features transform into the fourfold pattern of risk attitudes in quantity perception: for low probabilities, individuals are risk seeking in the domain of gains but risk averse in the domain of losses; for moderate and high probabilities, individuals are risk averse in the domain of gains but risk seeking in the domain of losses. In summary, the prospect theory formulates the non-linear quantity and probability perception of human psychology, which has profound implications for developing human-centric communications in both academia and industry. 

\section{Prospect Theoretic Analytical Framework}\label{ptbaf}
The basic mathematical formulation based on the prospect theory enables the mapping from objective QoS to subjective QoE, capturing the behavioral, psychological, and contextual factors. In the process of quantity and probability perception, the QoE or UE is dependent on two metrics: the quantity metric and the probability metric.

\subsection{Quantity Metric and Value Function}
The quantity metric denotes conventional QoS parameters such as bandwidth and latency. According to the prospect theory, individuals perceived value through state transition (i.e., quantity deviations from a reference point) instead of the current quantity. In various disciplines, utility/value functions serve as the standard approach for modeling UE. Following the two-part functional form of \cite{tversky1992advances}, the value function for telecommunications taking the reference dependence, diminishing marginal utility, and loss aversion into consideration can be modeled as\footnote{Note that, there is a prerequisite for using this value model, the quantity metric of interest must be a \textit{desirable} metric. A desirable metric is a metric that will be preferable with a larger value, e.g., transmission rate and network coverage. In this paper, we only consider  desirable metrics without special notes, because of the limitation of the prospect theory originally dealing with the monetary benefit that is also a desirable metric.}$^{,}$\footnote{For simplicity, we mainly analyze the single-metric scenario, in which only a single quantity metric $x$ with its reference point $x_0$ is taken into consideration.}
\begin{equation}\label{valuefunc}
v(x,x_0)=\begin{cases}
\lambda_1(x-x_0)^{\alpha_1},~~~~~~x\geq x_0\\
-\lambda_2(x_0-x)^{\alpha_2},~~~~x< x_0,\\
\end{cases}
\end{equation}
where $\alpha_1$, $\alpha_2$, $\lambda_1$, and $\lambda_2$ are positive and user specific parameters;  $x_0>0$ is the reference point with respect to $x$, which captures the reference dependence. The reference point $x_0$ can be a previous quantity, expected quantity, or contractual quantity for different application scenarios. For generality, we model the process of quantity perception as a function of both $x$ and $x_0$ so as to emphasize the equal importance of the actual quantity $x$ and the reference point $x_0$. Both dimensions are indispensable in determining user perception of quantity metrics. It is worth noting that all the parameters, including the reference point $x_0$ are specific in terms of user preferences and socio-economic contexts and could change over time. Obviously, if $\alpha_1=\alpha_2=1$, $\lambda_1=\lambda_2=1$, and $x_0=0$, we have $v(x,x_0)=x$, $\forall~x\geq 0$, and the formulated analytical framework reduces to the classic QoS analytical framework.

Without loss of generality, the value model given in (\ref{valuefunc}) is called the four-parameter value model, which is different from the classic two-parameter value model widely used in behavioral economics that assumes $\alpha_1=\alpha_2$ and fixes $\lambda_1=1$. For the four-parameter value model, it is worth inspecting and discussing the constraints on parameters that should jointly ensure the key attributes retrieved from the prospect theory. It has been summarized in \cite{dhami2016foundations} that any value function $v(x,x_0)$ complying with the prospect theory must satisfy the following fundamental properties:
\begin{itemize}
\item $v(x,x_0)$ is continuous and strictly increasing with respect to $x$;
\item $v(x_0,x_0)=0$ (reference dependence);
\item With respect to $x$, $v(x,x_0)$ is concave when $x\geq x_0$ and is convex when $x<x_0$ (diminishing marginal utility);
\item $v(x_0+\delta,x_0)<-v(x_0-\delta,x_0)$, $\forall~\delta>0$ (loss aversion).
\end{itemize}
The first two fundamental properties of $v(x,x_0)$ are axiomatic for the four-parameter value model. To investigate the concavity and convexity, it is straightforward to derive the second-order piecewise partial derivative of $v(x,x_0)$ with respect to $x\geq x_0$ as $\frac{\partial^2v(x,x_0)}{\partial x^2}\vert_{x\geq x_0}=\alpha_1(\alpha_1-1)\lambda_1(x-x_0)^{\alpha_1-2}$ and $x<x_0$ as $\frac{\partial^2v(x,x_0)}{\partial x^2}\vert_{x< x_0}=-\alpha_2(\alpha_2-1)\lambda_2(x_0-x)^{\alpha_2-2}$. Therefore, solving $\frac{\partial^2v(x,x_0)}{\partial x^2}\vert_{x\geq x_0}<0$ and $\frac{\partial^2v(x,x_0)}{\partial x^2}\vert_{x< x_0}>0$ yields $0<\alpha_1<1$ and $0<\alpha_2<1$, respectively. For the last property stipulating $v(x_0+\delta,x_0)<-v(x_0-\delta,x_0)$, $\forall~\delta>0$, we can simplify the inequality to $\frac{\lambda_2}{\lambda_1}\delta^{\alpha_2-\alpha_1}>1$, $\forall~\delta>0$. By rigorous analysis, it can be proven that the only approach to ensure the validity of the inequality regardless of the value of $\delta$ is to let $\alpha_1=\alpha_2$ and $\lambda_1<\lambda_2$, which reduce the four-parameter model constructed in (\ref{valuefunc}) to a three-parameter model regulating $\alpha=\alpha_1=\alpha_2\in(0,1)$ and $\lambda_1<\lambda_2$. Rigorous and comprehensive discussions and proofs regarding the parameters of value model in the prospect theory can be found in \cite{al2008note}. 

\begin{figure*}
    
    \begin{subfigure}[t]{0.3\textwidth}
        \includegraphics[width=2.5in]{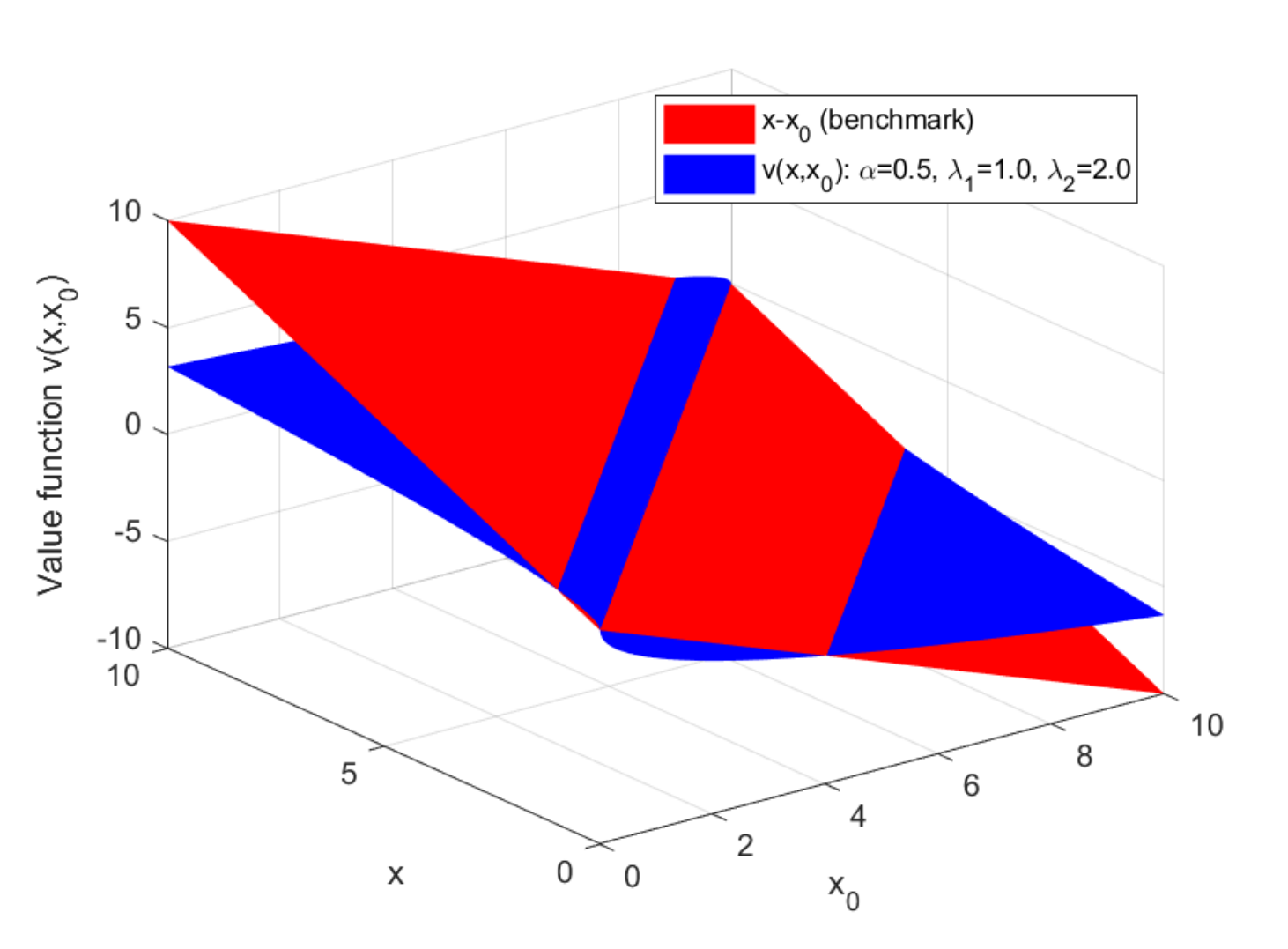}
        \caption{}
        \label{}
    \end{subfigure}
    ~~ 
    \begin{subfigure}[t]{0.3\textwidth}
        \includegraphics[width=2.5in]{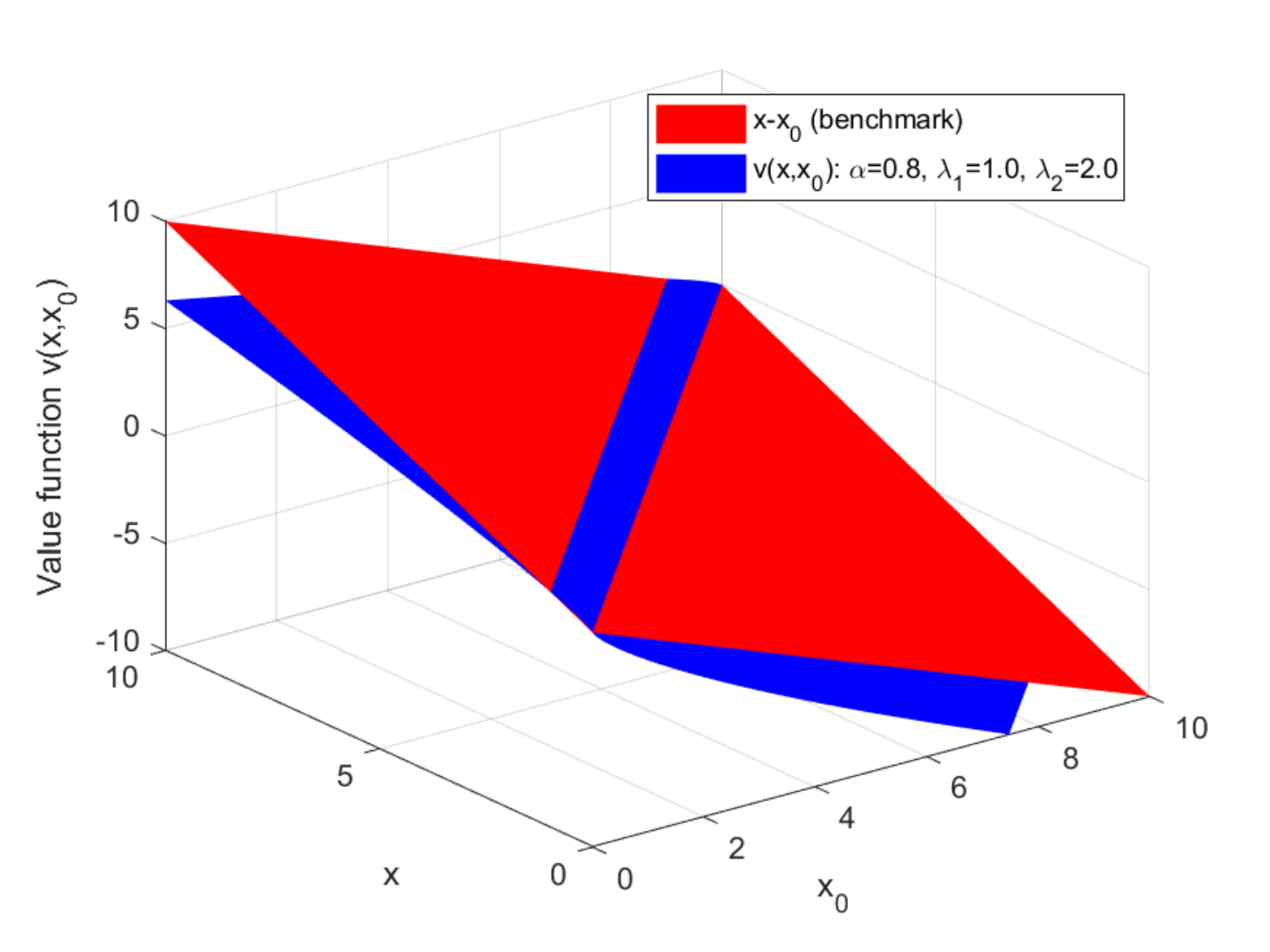}
        \caption{}
        \label{}
    \end{subfigure}
    ~~ 
    \begin{subfigure}[t]{0.3\textwidth}
        \includegraphics[width=2.5in]{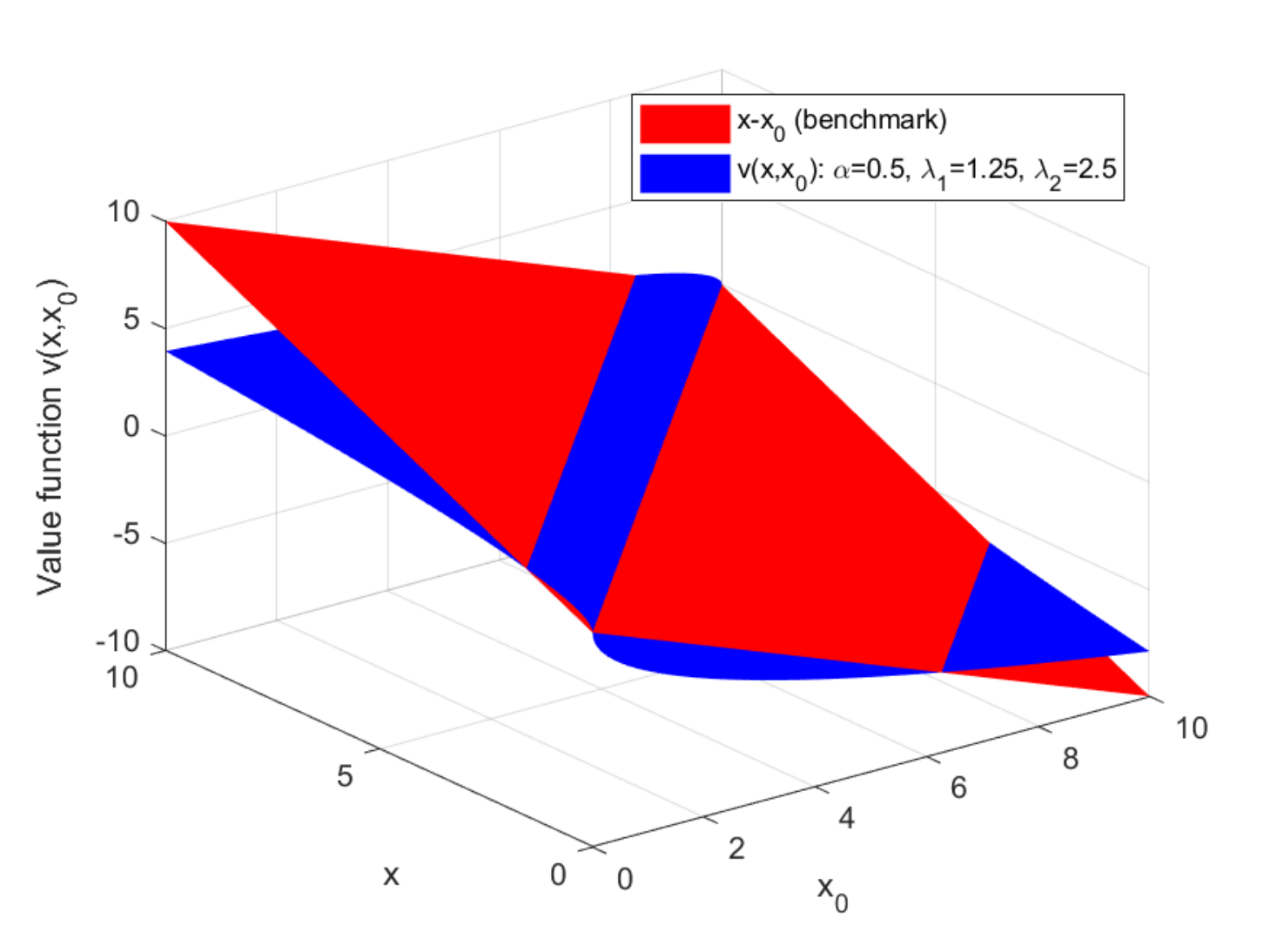}
        \caption{}
        \label{}
    \end{subfigure}
    \caption{Perceived values of state transitions with different sets of parameters. }
    \label{valuefuncplot}
\end{figure*}

\begin{figure}[!t]
\centering
\includegraphics[width=3.5in]{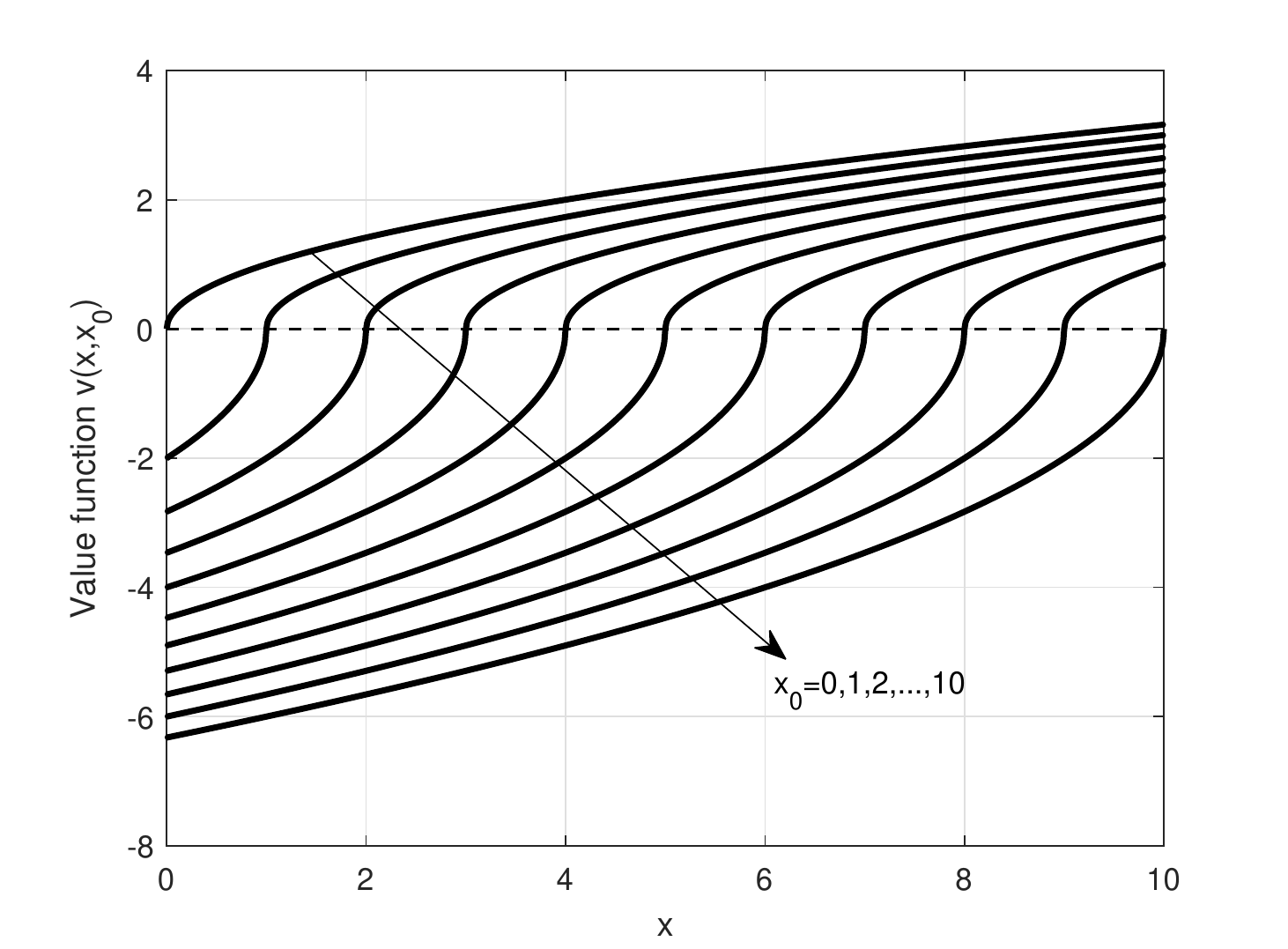}
\caption{Perceived value vs. quantity metric with different reference points, given $\alpha=0.5$, $\lambda_1=1.0$, and $\lambda_2=2.0$.}
\label{referenceplot}
\end{figure}

For illustration purposes, we plot $v(x,x_0)$ with different sets of parameters in Fig. \ref{valuefuncplot} by referring to the suggested parameter ranges yielded by empirical evidence given in \cite{dhami2016foundations}. We also plot Fig. \ref{referenceplot} to illustrate the effects of reference point on the value function. In these figures, we can confirm that the key attributes related to the value function from the prospect theory hold and inspect how these key attributes and the relevant parameters jointly affect the user perception of quantity metrics.

\subsection{Probability Metric and Probability Weighting Function}
The probability metric in the context of telecommunications denotes the measurement of opportunistic performance, which encompasses outage probability, error probability, collision probability, handover probability, and etc. An objective probability will be distorted when being perceived by end users, corresponding to the last key attribute of the prospect theory: non-linear probability perception. Introducing the Prelec probability weighting function \cite{prelec1998probability}, we model the perception of a probability metric $0\leq p\leq 1$ under psychological distortion as
\begin{equation}\label{vpxeq}
w(p)=\mathrm{exp}(-\gamma(-\log(p))^\theta),
\end{equation}
where $\gamma>0$ and $0<\theta<1$ are user specific parameters used to characterize the subjective perception of probability metric. The ranges $\gamma$ and $\theta$ are derived by the necessary conditions for maintaining the basic properties of probability weighting function, especially its inverse S-shape. Detailed discussions on the functionality and property of $\gamma$ and $\theta$ can be found in \cite{dhami2016foundations}. If $\gamma=1$ and $\theta=1$, we have $w(p)=p$, $\forall~p\in[0,1]$, and thus the formulated analytical framework for probability perception reduces to the classic QoS analytical framework. 

In a general case with a continuous random variable $X$, the probability argument in (\ref{vpxeq}) could be any cumulative distribution function (CDF) $F_X(s)=\mathbb{P}\{X\leq s\}$, where $\mathbb{P}\{\cdot\}$ denotes the objective probability of the random event enclosed. Following the definition of CDF and the attribute of probability distortion, we gives the perceptual CDF (PCDF) by
\begin{equation}
\tilde{F}_X(s)=w(F_X(s)).
\end{equation}
Denoting $f_X(s)=\frac{\mathrm{d}F(s)}{\mathrm{d}s}$ as the probability density function (PDF) of $X$, we can define the perceptual PDF (PPDF) as 
\begin{equation}
\begin{split}
\tilde{f}_X(s)&=\frac{\mathrm{d}\tilde{F}_X(s)}{\mathrm{d}s}=\frac{\mathrm{d}w(F_X(s))}{\mathrm{d}s}\\
&=\gamma\theta w(F_X(s))(-\log(F_X(s)))^{\theta-1}f_X(s)/F_X(s).
\end{split}
\end{equation}

\begin{figure*}
    
    \begin{subfigure}[t]{0.3\textwidth}
        \includegraphics[width=2.5in]{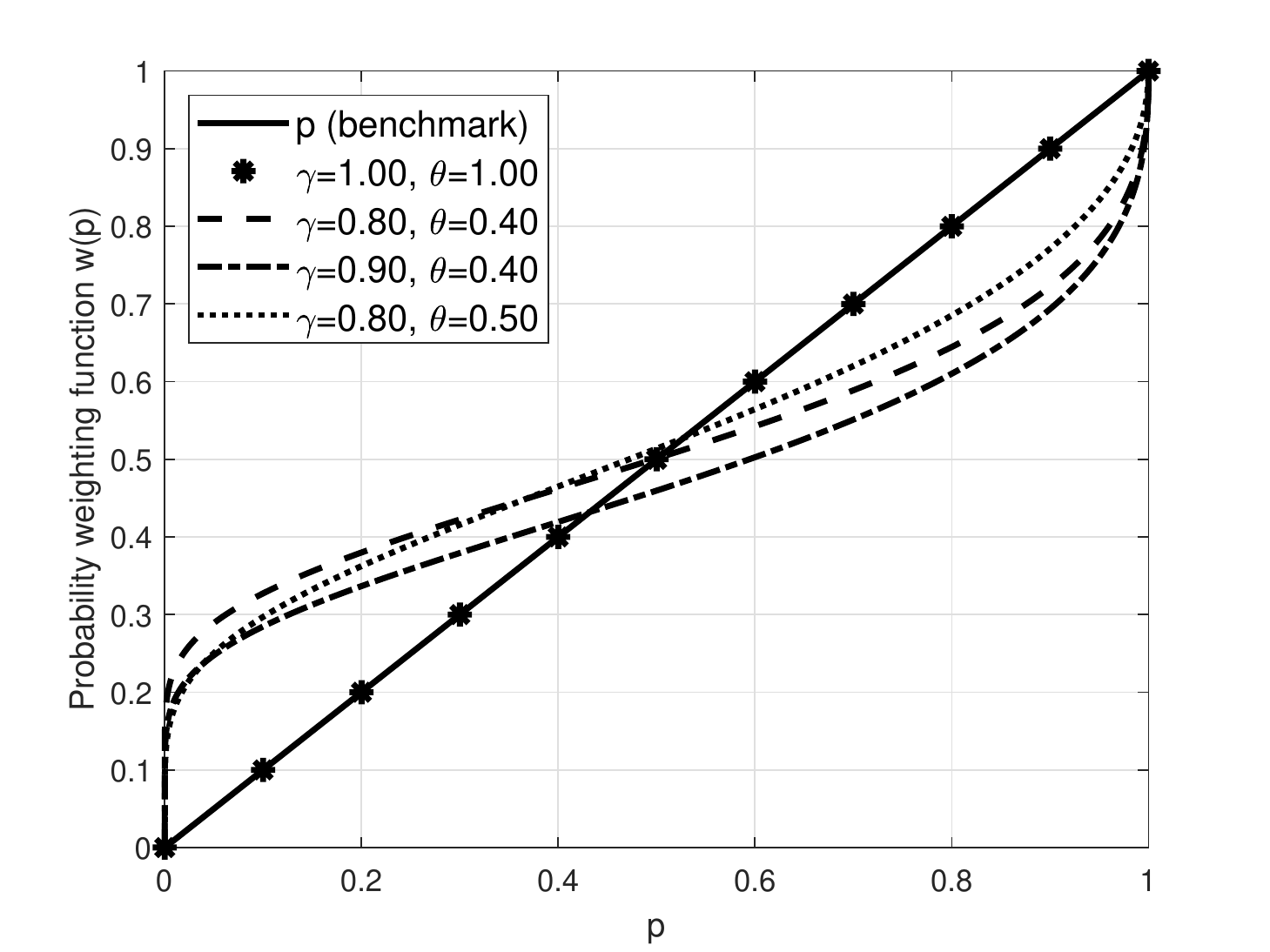}
        \caption{}
        \label{}
    \end{subfigure}
    ~~ 
    \begin{subfigure}[t]{0.3\textwidth}
        \includegraphics[width=2.5in]{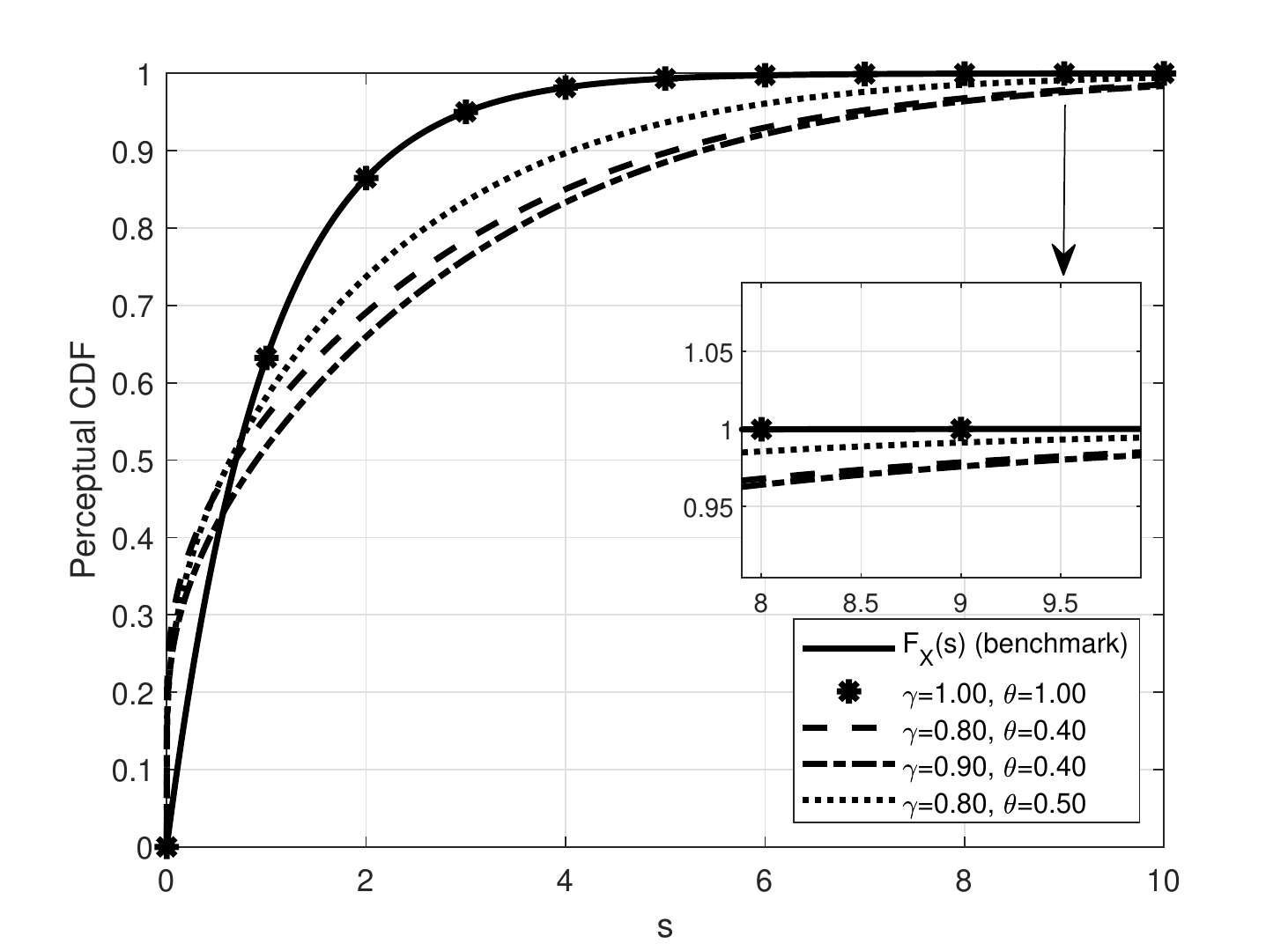}
        \caption{}
        \label{}
    \end{subfigure}
    ~~ 
    \begin{subfigure}[t]{0.3\textwidth}
        \includegraphics[width=2.5in]{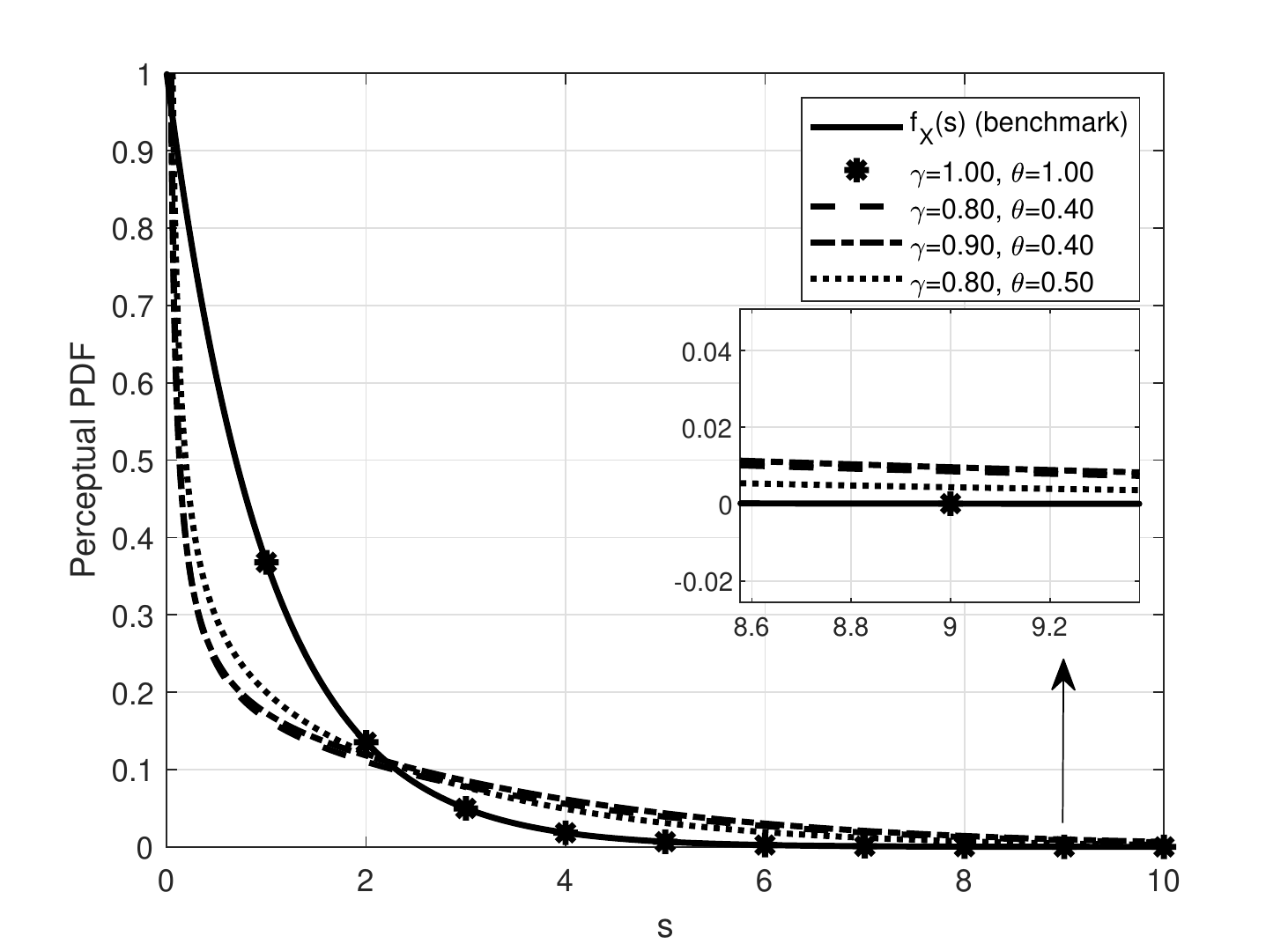}
        \caption{}
        \label{}
    \end{subfigure}
    \caption{Perceived probability, PCDF, and PPDF with different sets of parameters. Here, we take the normalized exponential distribution as an example to illustrate PCDF and PPDF, i.e., $F_X(s)=1-\mathrm{exp}(-s)$ and $f_X(s)=\mathrm{exp}(-s)$ for $s\geq0$.}
    \label{probfuncplot}
\end{figure*}

For illustration purposes, we plot $w(p)$, $\tilde{F}_X(s)$, and $\tilde{f}_X(s)$ in Fig. \ref{probfuncplot} by referring to the suggested parameter ranges yielded by empirical evidence given in \cite{dhami2016foundations}, ditto. In this figure, we can testify that the attribute of probability distortion from the prospect theory hold and inspect how this attribute and the parameters $\gamma$ and $\theta$ jointly affect the user perception of probability metrics. 

\subsection{Summary of the Proposed Analytical Framework}
\begin{figure}[!t]
\centering
\includegraphics[width=3.5in]{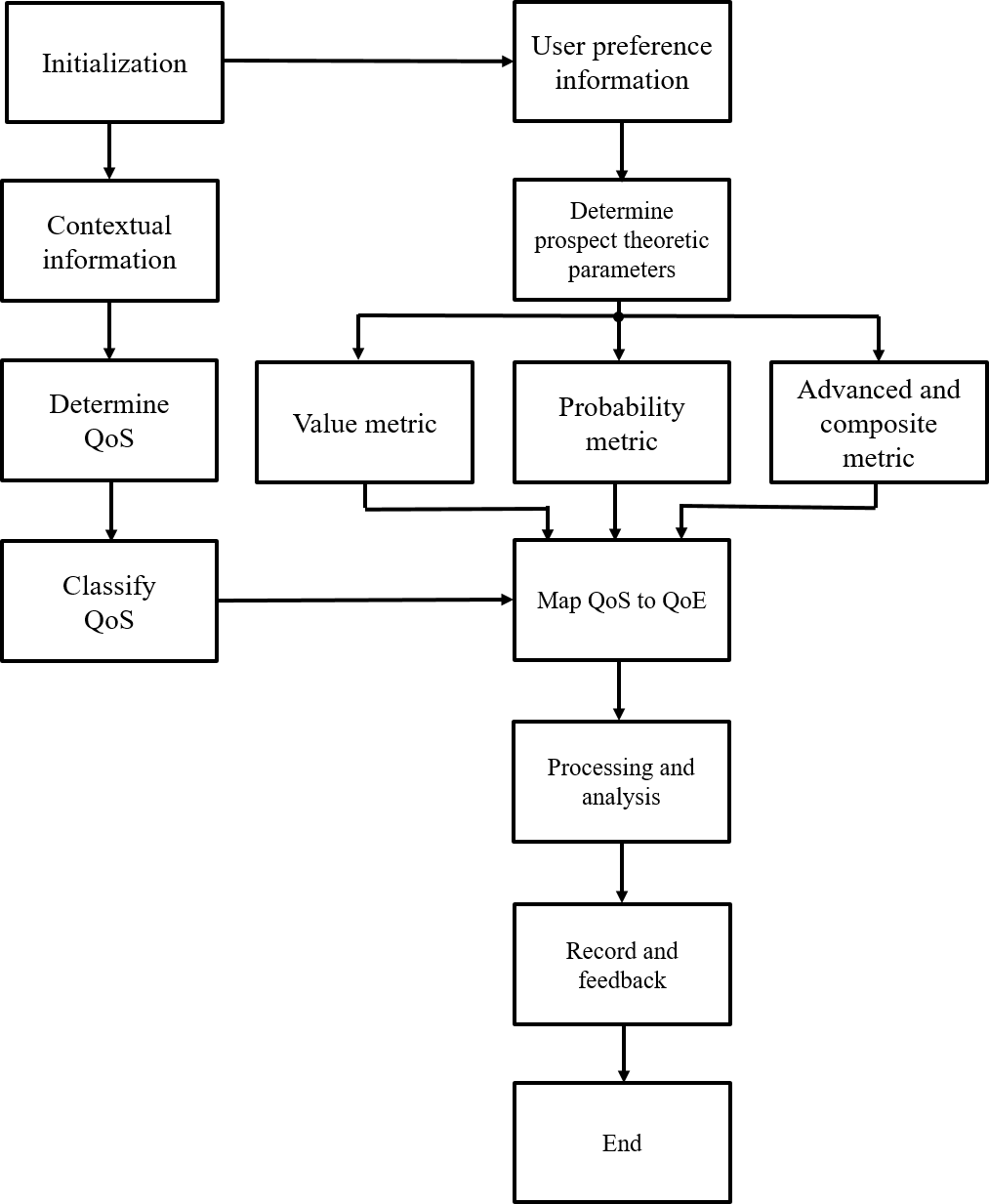}
\caption{Implementation process of the prospect theoretic analytical framework for perceptual performance analysis.}
\label{implementationsys}
\end{figure}

Having obtained $v(x,x_0)$, $w(p)$, $\tilde{F}_X(s)$, and $\tilde{f}_X(s)$, we can reevaluate a set of advanced and composite performance metrics for communication systems incorporating non-technological factors. For example, given a composite metric $\Omega(g)$ that is a function of a random variable $g$, and the PDF of $g$ is denoted as $f_G(g)$, we can define the perceptual utility (PU) of the composite metric by $\tilde{\Omega}=\int_{0}^{\infty}v(\Omega(g),\Omega_0)\tilde{f}_G(g)\mathrm{d}g$, where $\Omega_0$ is the reference point of the composite metric. $\tilde{\Omega}$ can be employed to appraise the subjective performance pertaining to the composite variable. Different from objective performance evaluation metrics, the PU based on the prospect theory is allowed to be negative, which implies a negative user impression/perception of the objective performance provided. Fig. \ref{implementationsys} depicts the complete implementation process of the prospect theoretic analytical framework for perceptual performance analysis. Specific case studies for applying the analytical framework are given in the next section.

\section{Case Studies}\label{cs}
\begin{figure*}[!t]
\centering
\includegraphics[width=5.5in]{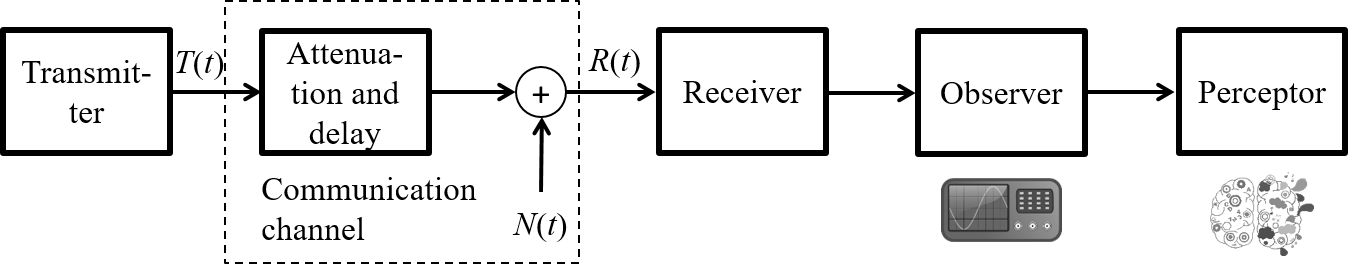}
\caption{Communication system model considered in this paper.}
\label{sys}
\end{figure*}

Consider a simplistic point-to-point (P2P) wireless communication system consisting of a transmitter, a receiver, and a time-variant fading channel obeying the Rayleigh distribution, in which received signals are attenuated by multi-path fading and contaminated by a complex additive white Gaussian noise (AWGN). However, different from the classic communication system model, we add an \textit{observer} and a \textit{perceptor} after the receiving module, as shown in Fig. \ref{sys}\footnote{Note that, the proposed communication system model with the perceptor is different from the user-in-the-loop (UIL) model given in \cite{6736762}, which relies on incentives (e.g., dynamic pricing) to change user behaviors and responses.}. The observer is responsible for evaluating received signals on an objective basis, so that the QoS metrics can be retrieved, whereas the perceptor is employed to assess the QoS metrics on a subjective basis to obtain the perceptual metrics.

In such a simplistic P2P wireless communication system, the equivalent baseband input-output relation in the time domain can be written as \cite{proakis2008digital}
\begin{equation}\label{redasdas}
\begin{split}
&R(t)\\
&=\sqrt{P_t}\sum_{k=1}^{K}\left(A_k(t)\mathrm{exp}(-j2\pi f_c \tau_k(t))T(t-\tau_k(t))\right)+N(t),
\end{split}
\end{equation}
where $T(t)$ and $R(t)$ represent the baseband transmitted and received signals at time $t$, respectively; $K$ is the number of propagation paths yielded by direct propagation, reflection, refraction, and scattering; $A_k(t)$ and $\tau_k(t)$ characterize the attenuation and delay of the $k$th propagation path; $f_c$ is the central carrier frequency of the transmitted signal; $P_t$ is the transmit power; $N(t)$ represents the AWGN sample at the receiver with an average noise power $N_0$. In telecommunications, $\tau_k(t)$ is generally assumed to be an exponentially distributed random variable and $f_c\gg 1$. As a result, $\theta_k(t)=2\pi f_c\tau_k(t)$ is approximated to be a uniform distributed random variable between 0 to $2\pi$ rad. When the transmission is within a signaling interval, we can write the channel coefficient as
\begin{equation}
H(t)=\sum_{k=1}^{K}A_k(t)\mathrm{exp}(-j\theta_k(t))),
\end{equation}
where $H(t)$ is a zero-mean complex Gaussian distributed random variable according to the central limit theorem when $K\rightarrow\infty$. For simplicity, we can assume that $G(t)=|H(t)|^2$ is exponentially distributed with CDF $F_G(g)=1-\mathrm{exp}(-g/\mu)$ and PDF $f_G(g)=\mathrm{exp}(-g/\mu)/\mu$ for $g\geq 0$, where $\mu$ is the average channel power gain \cite{8344837}.

Based on this simplistic model, we analyze the perceptual metrics corresponding to three fundamental but important QoS metrics: signal-to-noise ratio (SNR), transmission rate, and outage probability as follows, through the prospect theoretic analytical framework built in the last section.

\subsection{PU of Signal-to-Noise Ratio}
\begin{figure}[!t]
\centering
\includegraphics[width=3.5in]{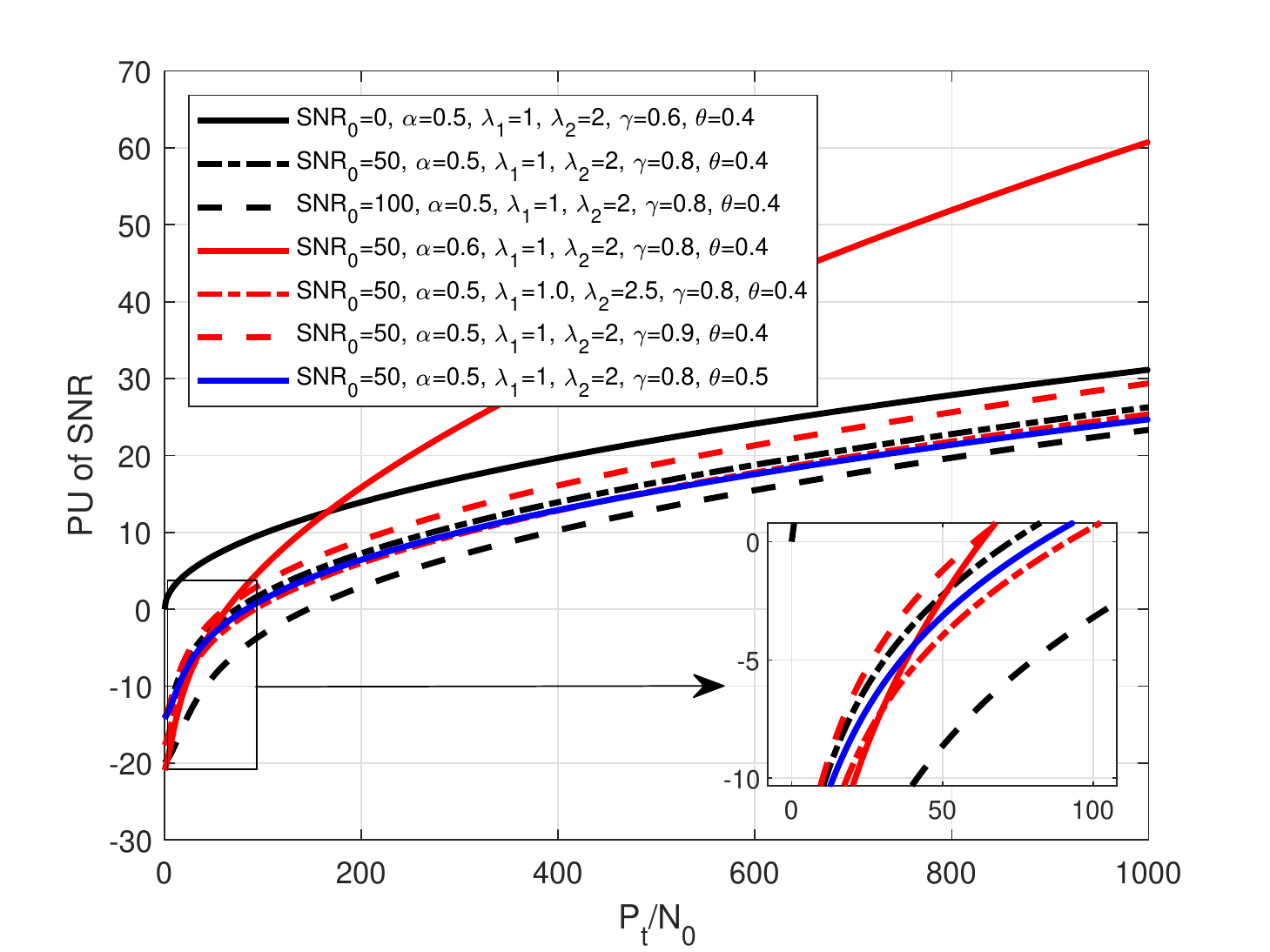}
\caption{PU of SNR with different sets of parameters, given $\mu=1$.}
\label{pusnr}
\end{figure}

First of all, the instantaneous SNR can be expressed as
\begin{equation}\label{instantsnrexp}
\Gamma(t)={P_tG(t)}/{N_0}.
\end{equation}
To analyze the average perception of continuous outcomes instead of discrete values, researchers in \cite{rieger2006cumulative} generalize the original formulation of utility given in \cite{tversky1992advances} and provide a new utility metric termed the \textit{subjective utility}. Tailoring the subjective utility proposed by \cite{rieger2006cumulative}, we define the PU of SNR in the context of the prospect theory as follows:
\begin{definition}
\textit{The prospect theoretic PU of SNR is the average perceived value of SNR from the user's perspective, complying with the human psychology of non-linear quantity and probability perception.}
\end{definition}
With the proposed analytical framework and the definition given above, the prospect theoretic PU of SNR can be written as 
\begin{equation}\label{snreq}
\begin{split}
\widetilde{\Gamma}&=\int_{0}^{\infty} v(\Gamma(t),\Gamma_0)\tilde{f}_G(G(t))\mathrm{d}G(t)\\
&=-\lambda_2\int_{0}^{\frac{N_0\Gamma_0}{P_t}} \left(\Gamma_0-\Gamma(t)\right)^{\alpha}\tilde{f}_G(G(t))\mathrm{d}G(t)\\
&~~~~+\lambda_1\int_{\frac{N_0\Gamma_0}{P_t}}^{\infty} \left(\Gamma(t)-\Gamma_0\right)^{\alpha}\tilde{f}_G(G(t))\mathrm{d}G(t),
\end{split}
\end{equation}
where $\Gamma_0$ is a predetermined reference point of SNR. The PU of SNR can be used to characterize the QoE/UE regarding the reliability of communication networks. For illustration purposes, we plot the PU of SNR versus $P_t/N_0$ with different sets of parameters for the normalized channel configuration (i.e., $\mu=1$) in Fig. \ref{pusnr}.

\subsection{PU of Transmission Rate}
\begin{figure}[!t]
\centering
\includegraphics[width=3.5in]{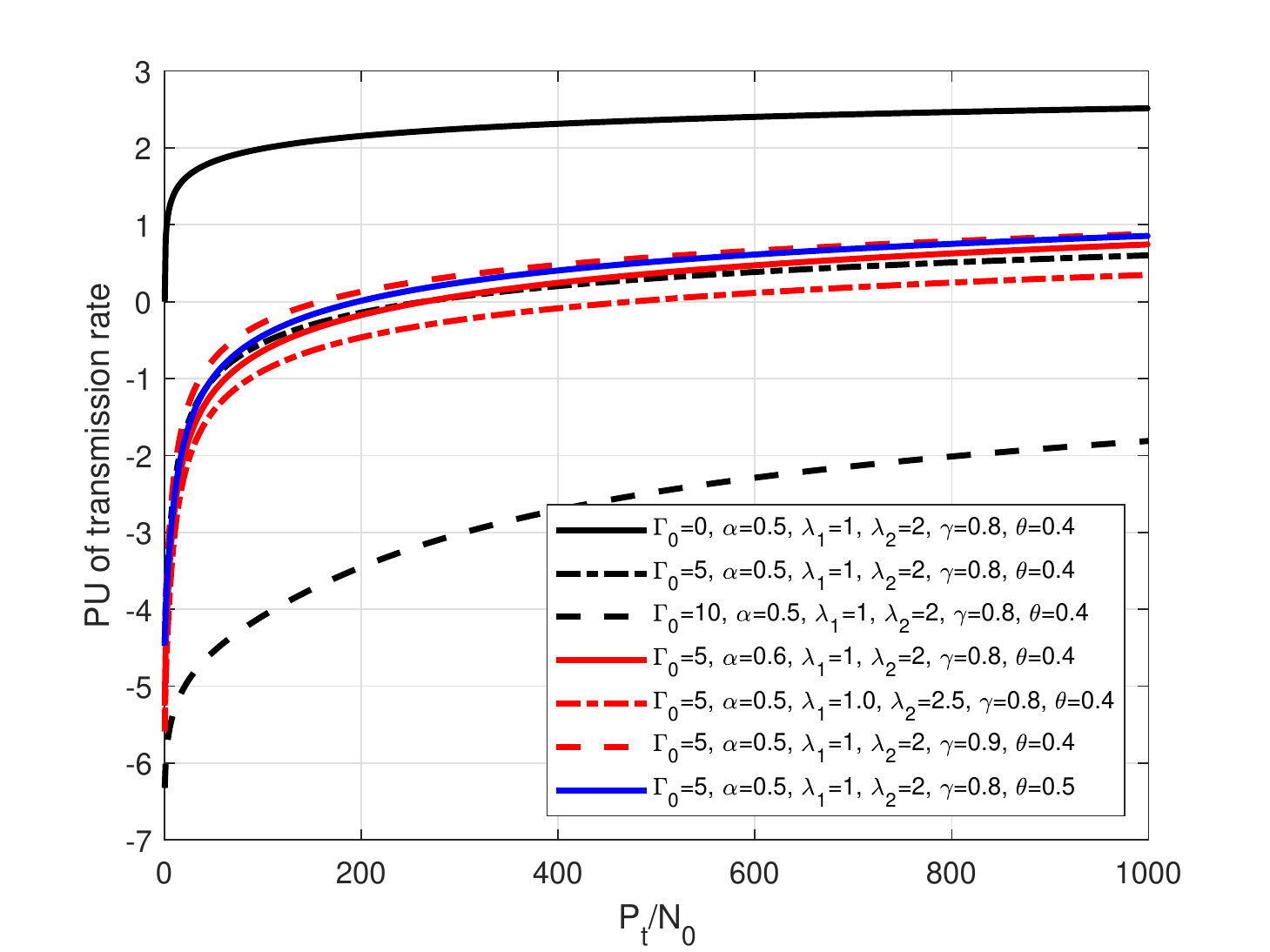}
\caption{PU of transmission rate with different sets of parameters, given $\mu=1$.}
\label{putr}
\end{figure}

Based on the formulation of the instantaneous SNR given in (\ref{instantsnrexp}), the instantaneous transmission rate for the normalized bandwidth is given by \cite{7809043}
\begin{equation}
\Psi(t)=\log_2(1+\Gamma(t)).
\end{equation}
Similarly, according to the formulation of the subjective utility, we can define the PU of transmission rate in the context of the prospect theory infra:
\begin{definition}
\textit{The prospect theoretic PU of transmission rate is the average perceived value of transmission rate from the user's perspective, complying with the human psychology of non-linear quantity and probability perception.}
\end{definition}
The prospect theoretic PU of transmission rate can be explicitly expressed as
\begin{equation}\label{rateeq}
\begin{split}
\widetilde{\Psi}&=\int_{0}^{\infty} v(\Psi(t),\Psi_0)\tilde{f}_G(G(t))\mathrm{d}G(t)\\
&=-\lambda_2\int_{0}^{\frac{N_0\Psi_0}{P_t}} \left(\Psi_0-\Psi(t)\right)^{\alpha}\tilde{f}_G(G(t))\mathrm{d}G(t)\\
&~~~~+\lambda_1\int_{\frac{N_0\Psi_0}{P_t}}^{\infty} \left(\Psi(t)-\Psi_0\right)^{\alpha}\tilde{f}_G(G(t))\mathrm{d}G(t),
\end{split}
\end{equation}
where $\Psi_0$ is a predetermined reference point of transmission rate. The PU of transmission rate can be used to characterize the QoE/UE pertaining to the efficiency of communication networks. We plot the PU of transmission rate versus $P_t/N_0$ with different sets of parameters for normalized channel configuration in Fig. \ref{putr}.

\subsection{Perceptual Outage Probability}
\begin{figure}[!t]
\centering
\includegraphics[width=3.5in]{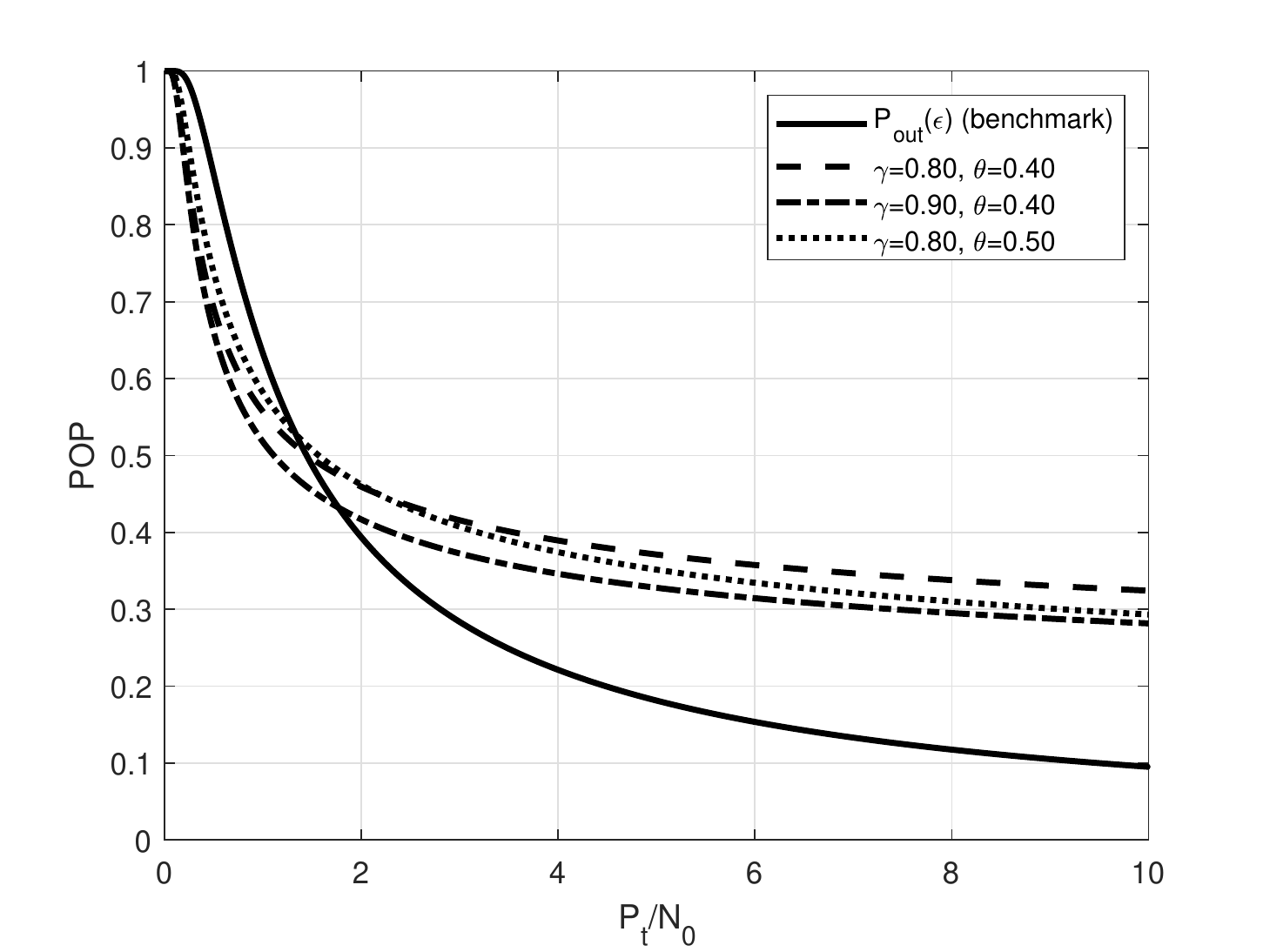}
\caption{POP with different sets of parameters, given $\mu=1$ and $\epsilon=1$.}
\label{pop}
\end{figure}

Following the formulation of the instantaneous transmission rate, we model the outage probability infra \cite{8703169}:
\begin{equation}
\begin{split}
P_{\mathrm{out}}(\epsilon)&=\mathbb{P}\left\lbrace\Psi(t)<\epsilon\right\rbrace\\
&=\mathbb{P}\left\lbrace \Gamma(t)<2^{\epsilon}-1\right\rbrace\\
&=\mathbb{P}\left\lbrace G(t)<N_0(2^{\epsilon}-1)/P_t\right\rbrace\\
&=F_G(N_0(2^{\epsilon}-1)/P_t),
\end{split}
\end{equation}
where $\epsilon$ is a predefined outage threshold related to communication system configurations. We can define the perceptual outage probability (POP) in the context of the prospect theory as
\begin{definition}
\textit{The POP is the perceived probability from the user's perspective, complying with the human psychology of non-linear probability weighting.}
\end{definition}
Therefore, we employ the Prelec probability weighting function to quantify the prospect theoretic POP as follows:
\begin{equation}
\begin{split}
\tilde{P}_{\mathrm{out}}(\epsilon)&=w(P_{\mathrm{out}}(\epsilon))=w(F_G(N_0(2^{\epsilon}-1)/P_t))\\
&=\tilde{F}_G(N_0(2^{\epsilon}-1)/P_t).
\end{split}
\end{equation}
The POP captures the distorted perception of outage probability and can be utilized to appraise the QoE/UE of outage. Normalizing parameters $\mu=1$ and $\epsilon=1$, we plot the POP versus $P_t/N_0$ with different sets of parameters in Fig. \ref{pop}. 

\subsection{Discussion}
According to the plotted data in Fig. \ref{pusnr}, Fig. \ref{putr}, and Fig. \ref{pop}, we can observe that the proposed perceptual metrics exhibit favorable mathematical properties, e.g., continuity, differentiability, and convexity/concavity. They also perform well in both of the low and high SNR regions, which are aligned with the prediction given by the prospect theory. Moreover, compared to other widely adopted value models, e.g., the logarithmic, sigmoid, and exponential value models \cite{reichl2013logarithmic,6891176,5430142}, the proposed value model relying on the power function is more computationally efficient. Fig. \ref{pusnr} and Fig. \ref{putr} have well captured the human psychology of reference dependence, diminishing sensitivity, and loss aversion; Fig. \ref{pop} demonstrates the phenomenon of subjective probability distortion. 

Overall, the plotted data is consistent with the model's economic implications. A smaller $\gamma$ denotes a wider range of over-weighted possibilities under exponential distribution, which leads to greater probability distortion and a lower prospect theoretic PU. For $0<\theta<1$, a smaller $\theta$ is associated with sharper curvature of the inverse S-shape and greater probability distortion. A higher reference point indicates `higher expectation' and leads to a lower or even negative QoE. A smaller $\alpha$ indicates a higher degree of diminishing utility and a lower overall utility. A larger $\lambda_2 / \lambda_1$ ratio indicates a higher degree of loss aversion and, thereby, a lower QoE. In summary, UE is significantly and negatively related to the psychological distortion in quantity and probability perception, consistent with the economic intuition that irrationality causes non-optimal outcomes. 

As an intriguing finding, Fig. \ref{pusnr} and Fig. \ref{putr} reveal that prospect theoretic users exhibit abnormally high degrees of loss aversion, reference dependence, and diminishing marginal utility in choosing communication services. Specifically, users with reasonable expectations (reference values) suffer a PU of -4 when $P_t/N_0$ approaches zero, whereas they enjoy an increase in PU of merely 0.3 when $P_t/N_0$ increases from 400 to 1000. This evidence is in line with the widespread scepticism that users may be insensitive to the exponential increase in communication performance \cite{kilkki2008quality}. It also lends credits to the core development strategy of 5G/6G networks that communications should focus more on user experience and stability improvement rather than a blind pursuit of performance metrics. 

The above analysis, though preliminary and tentative, has illustrated the importance of modeling users' psychological foundations in both QoE measurements and human-centric communications.

\section{Challenges and Promising Research Directions}\label{fird}
In this study, we introduce the prospect theoretic analytical framework for modeling UE in communication systems. Although preliminary and immature, the proposed framework  serves as a promising architecture for developing interdisciplinary applications of human-centric communications. In this section, we outline the challenges, potentials, and application aspects for future research on the prospect theoretic framework. 

\subsection{Extend Experimental Evidence to Non-Monetary Contexts}
Historically, the prospect theory has been developed based on monetary experiments (e.g., lottery-choice experiments). On the other hand, interdisciplinary applications of the prospect theory are becoming increasingly popular and have hitherto resulted in over a thousand research articles. In these applications, the analytical framework is based on the parameter estimates elicited by monetary experiments, for which an implicit assumption is that users must exhibit similar psychological features such as loss aversion in non-monetary contexts (e.g., the choice of communication services). However, this critical assumption may not hold. For instance, it is widely recognized that individuals have different quantity and probability preferences for monetary and non-monetary stimuli. As suggested by \cite{kilkki2008quality}, individuals seem less sensitive to QoS parameters (bit rates and delays) than monetary incentives, which indicates a higher degree of utility curvature. In the current stage, the practical significance of the QoE measurements remains unclear due to the lack of concrete empirical support \cite{saad2016toward}. There is an urgent need to extend the experimental evidence to non-monetary contexts, which can provide a solid foundation and an explicit guideline for future interdisciplinary applications of the prospect theory.  

\subsection{Upgrade the Proposed Framework Using Advanced Behavioral Theory}
In the past decade, the prospect theory has been successfully applied in diverse research topics that involve quantity and probability perception, which verifies the theory's potential in interdisciplinary applications. The prospect theory is now evolving at a rapid pace; the advances such as the third- and fourth-generation prospect theory are readily applicable and begin to attract researchers' attention. Beyond the prospect theory, there is a bigger picture for interdisciplinary applications of behavioral economics. For instance, behavioral game theory has the potential to reshape the research landscape for game theoretic communications (e.g., radio resource management). Behavioral time discounting has enriched our understanding of time preferences and has profound implications for the intertemporal allocation of network resources. Psychological foundations such as endowment effect, other-regarding preference, and judgment heuristics have promising application prospects in pricing strategy and product design in telecommunication industries. Overall, the paradigm shift from technology-centric to human-centric communications still faces tremendous challenges because communication science has long been preoccupied with classic QoS benchmarks. Human-centric communications remains a prospective uncultivated research area in the 5G era and the forthcoming 6G era.

\subsection{Develop Mathematical Tools}
Under the prospect theoretic framework, quantitative analysis of perpetual performance involves sophisticated mathematical manipulations (differentiation, integration, and power series expansion) on the value function $v(x,x_0)$ and the probability weighting function $w(p)$. However, the probability weighting function is generally employed within socioeconomics and lacks the application foundation in telecommunications. Due to the limited understanding of the mathematical properties of probability weighting function (especially the Prelec function), researchers always encounter challenges in solving the integral forms and find it difficult to further process the quantitative analysis \cite{rieger2006cumulative}. A complete application of the prospect theory in communication science requires further exploration of the mathematical properties and analytical tools for the proposed functional forms.

\subsection{Analyze the Information Theoretic Properties}
The information theory constructed by Claude E. Shannon in `A Mathematical Theory of Communication' has been widely acknowledged as a landmark in telecommunications and the mathematical foundation of modern communication systems \cite{shannon1948mathematical}. Developed from the information theoretic model, most communication system models consist of transmitter, channel, and receiver, in which objective noise, interference, impairments, and imperfection are taken into consideration \cite{8750780}. For developing human-centric communications, communication system models should incorporate the fundamental mechanism of human perception after the receiving module. We tentatively term this embodiment the \textit{perceptor} under the prospect theoretic framework. The introduction of \textit{perceptor} into communication systems can alter the formulation of information entropy and generate intriguing information theoretic properties.

\subsection{Analyze the Perceptual Performance of Advanced Applications}
In this paper, the prospect theoretic analytical framework was employed in several simplistic application scenarios for demonstration purposes. However, due to its generality and versatility, the proposed framework can be readily extended to advanced communication applications in beyond 5G networks. Promising application aspects include massive multiple-input and multiple-output (MIMO) systems, terahertz (THz) communication systems, visible light communication (VLC) systems, multi-user communication systems, cooperative relay aided communication systems, hybrid licensed/unlicensed communication systems, and etc. Overall, interdisciplinary collaborations between telecommunications, economics, and psychology is indispensable for the development of human-centric communications.

\section{Conclusion}\label{c}
In this paper, we proposed the prospect theoretic analytical framework for human-centric communications. This analytical framework takes non-technological factors and user psychology  into consideration when modeling and analyzing communication systems, which enables quantitative analysis of perceptual performance. We provided several applications in telecommunications to demonstrate how the proposed analytical framework can be employed to conduct performance analysis in a human-centric manner. We also outlined the unresolved problems and promising directions for future research. Overall, this paper provides a guideline for improving communication services and introduces a common platform that unifies the nomenclature and endeavor from different disciplines, including telecommunications, economics, and psychology. Meanwhile, we also aim to use this paper as a fuze to trigger a new interdisciplinary research area for further investigation.

\bibliographystyle{IEEEtran}
\bibliography{bib}

\end{document}